# Analysis of trends in human longevity by new model

Byung Mook Weon

LG.Philips Displays, 184, Gongdan1-dong, Gumi-city, GyungBuk, 730-702, South Korea

**Abstract**

Trends in human longevity are puzzling, especially when considering the limits of human longevity. Partially, the conflicting assertions are based upon demographic evidence and the interpretation of survival and mortality curves using the Gompertz model and the Weibull model; these models are sometimes considered to be incomplete in describing the entire curves. In this paper a new model is proposed to take the place of the traditional models. We directly analysed the rectangularity (the parts of the curves being shaped like a rectangle) of survival curves for 17 countries and for 1876-2001 in Switzerland (it being one of the longest-lived countries) with a new model. This model is derived from the Weibull survival function and is simply described by two parameters, in which the shape parameter indicates 'rectangularity' and characteristic life indicates the duration for survival to be '$\exp(-1) \approx 36.79\%$'. The shape parameter is essentially a function of age and it distinguishes humans from technical devices. We find that although characteristic life has increased up to the present time, the slope of the shape parameter for middle age has been saturated in recent decades and that the rectangularity above characteristic life has been suppressed, suggesting there are ultimate limits to human longevity. The new model and subsequent findings will contribute greatly to the interpretation and comprehension of our knowledge on the human ageing processes.

# 1. Introduction

The first explanatory model, and the most influential parametric mortality model in the literature, is that proposed by Benjamin Gompertz (Gompertz 1825). He recognised that an exponential pattern in age captured the behaviour of human mortality for large portions of the life table (Higgins 2003). Ever since Gompertz, many models have been suggested to mathematically describe survival and mortality curves (Thatcher et al. 1998), of which the Gompertz model (Gompertz 1825) and the Weibull model (Weibull 1951) are the most generally used at present (Gavrilov and Gavrilova 2001, 2003). Interestingly, the Gompertz model is more commonly used to describe biological systems, whereas the Weibull model is more commonly applicable to technical devices (Gavrilov and Gavrilova 2001, 2003). The survival curve for the Gompertz model has the form, $S=\exp[a/b(1-\exp(bt))]$, where $S$ is survival probability, $a$ and $b$ are constants (Wilmoth 1997). The survival curve for the Weibull model has the form, $S=\exp(-(t/\alpha)^{\beta})$, where $\alpha$ and $\beta$ are constants (Nelson 1990).

However, the Gompertz and Weibull models are limited to modelling the portion of survival and mortality curves associated with senescence (Higgins 2003). A better model is required to describe the whole portion of curve, to show detailed variations by ageing factors, and to stimulate scientists to create insight into the fundamental mechanisms of ageing. It is particularly the rectangularity of survival curves that is ubiquitous in recent trends of developed countries. The conventional interpretation of rectangularity seems to be only partially successful, and this may be due to the interpretation being based on inappropriate models (Eakin and Witten 1995).

It is recently discovered that human survival and mortality curves are well described by the following new mathematical model, derived from the Weibull survival function and simply described by two parameters, the age-dependent shape parameter and characteristic life,

$$S = \exp(-(t/\alpha)^{\beta(t)}) \qquad (1)$$

where $S$ denotes the survival probability of surviving to age $t$, $\alpha$ denotes characteristic life, and $\beta(t)$ denotes shape parameter as a function of age (see Materials and Methods about features of the two parameters). The original idea was obtained as follows: typical human survival curves show i) a rapid decline in survival in the first few years of life and ii) a relatively steady decline and then an abrupt decline near death thereafter (Fig. 1). Interestingly, the former behaviour resembles the Weibull survival function with $\beta<1$ and the latter behaviour seems to follow the case of $\beta >> 1$. With this in mind, it could be assumed that shape parameter is a function of age. The new model is completely different from the traditional Weibull model in terms of 'age-dependence of the shape parameter'. It is especially noted that the shape parameter '$\beta(t)=\ln(-\ln S)/\ln(t/\alpha)$' can indicate 'rectangularity' of the survival curve. The reason is that as the value of the shape parameter becomes a high value, the shape of the survival curve approaches a further rectangular shape. Mortality function is described by the mathematical relationship with the survival function (Gavrilov and Gavrilova 2001, 2003) and is as follows,

$$\mu = -\frac{d \ln S}{dt} = \frac{d}{dt}((t/\alpha)^{\beta(t)}) \qquad (2)$$

where $\mu$ is the mortality rate (or force of mortality), meaning the relative rate for the survival function decline (Gavrilov and Gavrilova 2001). Obviously, mortality trends should be directly associated with shape parameter trends. It is noteworthy that 'age-dependence of the shape parameter' intrinsically makes the mortality function inevitably complex and difficult for modelling (see Materials and Methods about the general form for the mortality function).

A fundamental question in ageing research is whether humans possess immutable limits to longevity (Wilmoth et al. 2000). Such a question has yet to be resolved, although more and more scientists are coming to believe that human longevity may be on the increase (Vaupel et al. 1998, Wilmoth et al. 2000, Oeppen and Vaupel 2002). For example, the two leading scientists, Olshansky and Austad, have made a bet on whether somebody would live past 150 years in 2150 (Science 2001). According to Hayflick, "if we are to increase human life expectancy beyond the fifteen-year limit that would result if today's leading causes of death were resolved, more attention must be paid to basic research on ageing" (Hayflick 2000). Furthermore, since 1840, record life expectancy has increased by 2.5 years per decades, but will this march to longevity continue for many more decades (Vaupel et al. 2003)? If there are increases and/or limitations in the trends of human longevity at present, then how should we analyse it? With the application of useful methodology, this report can support a deeper understanding of the recent trends in human longevity. That is, we will apply simple methods to directly evaluate rectangularity and characteristic life (as a measure of longevity) and substantial demographic evidences will be shown to interpret the recent trends in human longevity for 17 countries and for the years 1876-2001 in Switzerland.

## 2. Materials and Methods

*The original data*

The original human survival data were taken from "Human Mortality Database": Survival probability (*S*) is expressed as a fraction ($l_x/l_o$) of the number of survivors ($l_x$) out of 100,000 persons ($l_o$) in the original life tables. We used the "Human Mortality Database" University of California, Berkeley (USA) and the Max Planck Institute for Demographic Research (Germany) (the data was downloaded on 25 November 2003).

*Features of the two parameters*

Conveniently, the value of characteristic life (*α*) is always found at the duration for survival to be 'exp(-1)', this is known as the characteristic life. This feature gives the advantage of looking for the value of *α* simply by graphical analysis of the survival curve. In turn, with the observed value of *α*, we can plot 'rectangularity' with age by the mathematical equivalence of '*β*(*t*)=ln(-ln*S*)/ln(*t*/*α*)'. If *β*(*t*) is not constant with age, this obviously implies that '*β*(*t*) is a function of age'. On the other hand, *β*(*t*) mathematically approaches infinity as the age *t* approaches the value of *α* or the denominator 'ln(*t*/*α*)' approaches zero. This feature of *β*(*t*) can leave 'traces of *α*' in the plot of *β*(*t*), so we can observe variations of *β*(*t*) and *α* at once in the plot of the shape parameters. If *β*(*t*) (except for the mathematical singularity (traces of *α*)) can be expressed by an adequate mathematical function, the survival and mortality functions can be calculated by the mathematically expressed *β*(*t*).

*Calculation for the survival and mortality functions*

For the calculation of the survival and mortality functions, we use the following functions:

$$S = \exp(-(t/\alpha)^{\beta(t)})$$

$$\mu = (t/\alpha)^{\beta(t)} \times \left[\frac{\beta(t)}{t} + \ln(t/\alpha) \times \frac{d\beta(t)}{dt}\right]$$

Only two parameters, $\beta(t)$ and $\alpha$, determine the survival and mortality functions. In empirical practice, we could use a linear expression for the mature phase (middle age) and a quadratic expression for senescence phase (from characteristic life to maximum age): $\beta(t) = \beta_0 + \beta_1 t + \beta_2 t^2 + ...$, where the associated coefficients were determined by a regression analysis in the plot of shape parameter curve. And thus, the derivative of $\beta(t)$ was obtained as follows: $d\beta(t)/dt = \beta_1 + 2\beta_2 t + ...$, which is the important original finding in this study, that is, the shape parameter for humans is a function of age.

## 3. Results

We have analysed recent trends in human longevity for 17 countries and for the years 1876-2001 in Switzerland, which is one of the longest-lived countries. The original life tables have been taken from the latest updated life tables of 'Human Mortality Database'. We have chosen the data type of 'by year of death' or 'period' life tables for all sexes. There are two types of life tables, the cohort (or generation) life tables and the period (or current) life tables. The cohort life table presents the mortality experience of a particular birth cohort, while the period life table presents what would happen to a hypothetical (or synthetic) cohort if it experienced throughout its entire life the mortality conditions of a particular period in time. The period life table may thus be characterized as rendering a 'snap-shot' of current mortality experience, and this shows the long-range implications of a set of age-specific death that prevailed in a given year (Arias 2002).

Recent human survival curves for 17 countries are shown in Fig. 2A. As is well known, survival curves in many countries have approached a rectangular shape (Nusselder and Mackenbach 1996, Vaupel 1997, Azbel 1999, Lee 2003). This phenomenon is originally known as 'rectangularization' of survival curve (Fries 1980). The characteristic lives were observed graphically and the ranking of countries having higher value of $\alpha$ were decided as follows (Table 1): Japan, Switzerland, Sweden, Italy, Norway, Austria, Canada, Germany, Finland, Netherlands, USA, England and Wales, Denmark, Lithuania, Hungary, Bulgaria, and Russia. The corresponding shape parameter curves can be plotted by virtue of the observed values of $\alpha$ in Fig. 2B. Particularly, we chose 'Switzerland' as an example of the longest-lived country, because its shape parameter shows the highest level (Fig. 2B). In this manner, we analysed historical trends of survival and shape parameter curves for 1876-2001 in Switzerland (Fig. 3A, 3B), where they are plotted by life data of every ten years from 1880 to 2000, including 1876 and 2001.

From the above analyses of survival and shape parameter curves, we observed several important facts: i) the shape parameter is a function of age, ii) the shape parameter indicates the successful rectangularity of the survival curve, supporting the recent trends in the duration of healthy life, iii) the graphical differences in survival curves are so very clear in shape parameter curves, especially for senescence, and iv) very interestingly, the traces of $α$ tend to increase together with the level of shape parameter. The analysis of the shape parameter curve shows significant advantages for interpreting human ageing. The variation with age of the shape parameter for humans is very interesting. In Fig. 4, the exponential function is a special case of the Weibull model when $β=1$, which means a constant death (failure) rate (Nelson 1990). Most technical devices following the Weibull model have a constant value of $β$ between the values of 0.5 and 5.0 (Nelson 1990). However, interestingly, as seen for humans (Fig. 2B, 3B), '$β(t)$ is a function of age'. Particularly for ageing patterns, $β(t)$ distinguishes humans from technical devices. It seems to show the difference between humans and technical devices in terms of 'robustness' – biological systems keep working even though cells inevitably acquire some faulty genes (Ball 2002). The fundamental difference for robustness (reliability) between biological systems and technical devices is obvious (Gavrilov and Gavrilova 2001).

## 4. Discussion

Specifically, we find that the value of $α$ has been significantly increasing from 1876 to 2001 in Switzerland (Fig. 5A); what then is the driving force for such increase? We attribute the driving force of characteristic life to a positive intervention by the human efforts to improve medical, social, and/or economic conditions. Of course there are more fundamental causes to increase characteristic life, but an in-depth study is beyond the scope of this report. The trends and causes of increased characteristic life (when $S=\exp(-1) \approx 36.79\%$) may be identical with those of average life (when $S=50\%$), issued in many other literature.

The variations of rectangularity of survival curve seem to be somewhat complex. For more careful consideration of the variations of rectangularity, we classified into three phases the ages of 30 and 70: the developmental phase below 30, the mature phase (middle age) between 30-70, and senescence phase above characteristic life (~70).

i) First, during the developmental phase below 30, we observe a significant increase of the shape parameter for 1876-2001 in Switzerland (Fig. 3B). Humans from conception to adulthood are virtually identical in their biological development (Hayflick 2000). Thus, developmental mechanisms may have a clear impact on ageing and are associated with a genomic impact. Furthermore, early-life conditions can determine the initial damage load acquired during early development (Gavrilov and Gavrilova 2003). The historical increase is attributable to the substantial elimination of infectious diseases that occur in youth through better hygiene and through the discovery of antibiotics and vaccines (Hayflick 2000).

ii) Second, during mature phase (middle age) between 30-70, we find that a linear expression for $\beta(t)$ is appropriate. Although characteristic life has increased continuously (Fig. 5A), the slope of the shape parameter has been saturated in recent decades (since about 1940) in Switzerland (Fig. 5B, and also see Fig. S1 in Supplementary file). The maximized slope of $\beta(t)$ is not beyond '0.1' on the average for both male and female. We note that continuous increase of characteristic life can lead to increases of the intercept of the shape parameter, regardless of the saturated slope of the shape parameter. Therefore, increase of rectangularity seems to be due to the increase of characteristic life. We think that the saturated slope of the shape parameter is important for interpreting modern human ageing, since it may suggest a possible limit to human longevity. Specifically, it may support the 'limited-distribution hypothesis' of all the possible hypotheses about limits to human longevity (Wilmoth 1997) – there exists a limiting distribution that mortality curves may approach but not surpass. After biological developments, death results from the inevitable increase in systemic molecular disorder (Hayflick 2000): "Just as a blueprint is vital for manufacturing a complex machine but contains no information to cause ageing, the genome is vital for biological development but contains no instruction for ageing". Thus, the reliability theory of ageing will be crucial (Gavrilov and Gavrilova 2001, 2003) and the systemic approach may be useful (Promislow and Pletcher 2002).

iii) Third, during the senescence phase after characteristic life to maximum age, we see that the trajectory of $β(t)$ or the 'rectangularity' has been suppressed for 1876-2001 in Switzerland (Fig. 6A). In this paper, the suppression indicates that the interval between characteristic life and maximum age has become shorter (Fig. 6A). We note that maximum age has increased much more slowly than the average life (Wilmoth et al. 2000) or characteristic life (Fig. 6A). Thus, such suppression may suggest a possible limit to human longevity. Of the total increase of lifespan, more than 70% is attributable to a decline in mortality above age 70 (Hayflick 2000, Wilmoth et al. 2000). Such survival is due overwhelmingly to improved medical practice concerning heart disease, stroke, smoking cessation, and the development of new drugs (Wilmoth et al. 2000). On the other hand, we estimated several trends of $β(t)$ (solid lines in Fig. 6A) for 1876, 1930, 1950, and 2001 (these years being arbitrarily selected for analysis), assuming that $β(t)$ has a quadratic relationship with age. In turn, we calculated the corresponding mortality curves from the estimated $β(t)$ (Fig. 6B). Interestingly, mortality deceleration seems to be directly associated with the 'bending' of the shape parameter by the suppression (Fig. 6B). The mortality deceleration is one of the central questions of demographers and biologists (Vaupel 1997, Higgins 2003). If the 'bending' is due to the suppression, the mortality deceleration would be a consequent phenomenon.

As a result, the findings suggest that there coexist promotion and limitation trends in human longevity at present. Nonetheless, it is expectable that for the time being, continuing progress of human society in advancing public health and the biomedical sciences will contribute to even longer and healthier lives. However, although characteristic life has increased up to date, the slope of the shape parameter for middle age has been saturated in recent decades and rectangularity for senescence above characteristic life has been suppressed, suggesting ultimate limits to human longevity, as shown in Fig. 7. If maximum age has increased much more slowly than average or characteristic life, there may be a concern for suppression, which would be dependent upon the increase rate of maximum age (unfortunately the slope of the increase rate is not clear in this study). It is noteworthy that the saturated slope of the shape parameter or 'rectangularity' in recent decades strongly supports the statement of the fifty-one leading scientists who study ageing (Olshansky et al. 2002a, 2002b): "the primary goal of biomedical research and efforts to slow ageing should not be the mere extension of life, but to prolong the duration of healthy life". Thus, to surpass the saturated slope of shape parameter would be a guide for the next strategy to prolong the duration of healthy human life. On the other hand, it is interesting that the mortality deceleration seems to be directly associated with the 'bending' of the shape parameter by the suppression.

## 5. Conclusions

In conclusion, the findings could support the assertion of limits of human longevity (Fig. 7). There are two factors of the limitation trends: i) the saturated slope of the shape parameter and ii) the suppression between characteristic (or average) life and maximum age. The second trend (the suppression) may be arguable, because the increase rate of characteristic (or average) life should be higher than that of maximum age. But, according to the previous literature (Wilmoth et al. 2000), it may be true that the suppression has occurred in recent decades. The first trend (the saturated slope) seems to be obvious in this study. We think that this is associated with Hayflick's point of view, "ageing is not a disease" (Hayflick 2000). The reliability theory (or system biology) is important to understand ageing processes (Gavrilov and Gavrilova 2001, 2003). The age-dependent shape parameter may be a key for the understanding of reliability behaviour. We are now showing an interesting observation and question: "the shape parameter is a function of age for humans (biological systems), whereas it is constant for technical systems, but why?" – we attribute this difference to the nature of biological systems, for example, self-repair or adaptation. Finally, we expect that the new model and subsequent findings will contribute greatly to the interpretation and comprehension of our knowledge on the human ageing processes.

**Acknowledgements**

The author is grateful to Dr. J. W. Vaupel and Dr. S. J. Olshansky for stimulating comments, and also to Dr. J. H. Je for useful discussions.

**Figure legends**

Fig. 1. An illustration of the original idea of the new model for typical human survival curve. It could be assumed that the shape parameter ($β$) of the traditional Weibull model would be a function of age.

Fig. 2. Survival and shape parameter curves for 17 countries. Recent human survival curves (A) and their shape parameter curves (B) Shape parameter curve of Switzerland shows the top level. Traces of characteristic life ($α$) are shown.

Fig. 3. Survival and shape parameter curves for 1876-2001 in Switzerland. Historical trends of human survival curves (A) and their shape parameter curves (B). Note that the shape parameter curve has age-dependence and shows good rectangularity.

Fig. 4. Schematic view of the concept of age-dependence of shape parameter. Shape parameter ($β$) is a function of age for humans, whereas it is just constant for technical devices.

Fig. 5. Historical trends for middle age (30-70). Historical trends of characteristic life (A) and slope of the shape parameter (B) for 1876-2001 in Switzerland. Although characteristic life has been continuously increasing, the slope of the shape parameter has been saturated in recent decades (since about 1940).

Fig. 6. Historical trends for senescence phase (above characteristic life). Historical trends of selected shape parameter curves (solid line in A) and calculated mortality curves (B) (plotted on a semi-log scale) for 1876-2001 in Switzerland. Solid lines in shape parameter behaviour indicate trend lines estimated by regression analysis. Mortality deceleration seems to be directly associated with 'bending' of the shape parameter by suppression (the interval between characteristic life and maximum age becomes shorter).

Fig. 7. Schematic view of trends in human longevity. Although characteristic life increases, the saturated slope of the shape parameter and suppression may stop the increase of longevity.

**Figures**

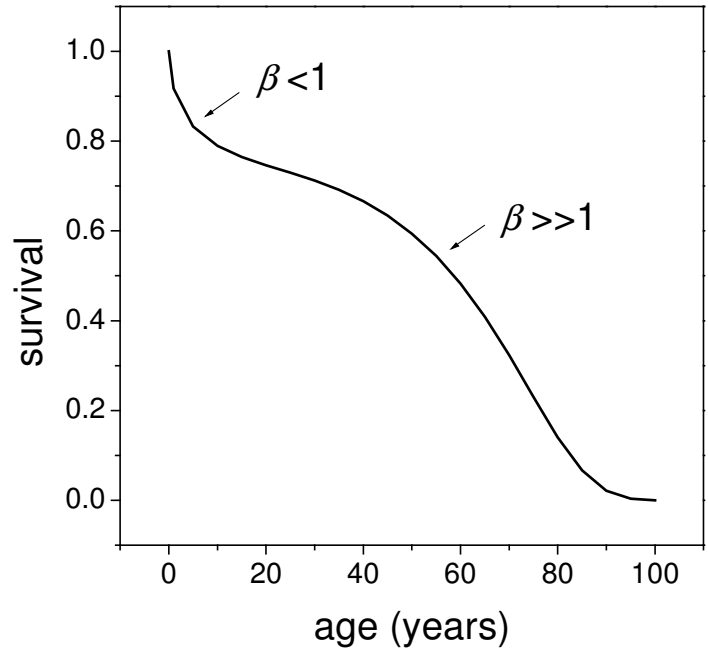

Fig. 1. Weon

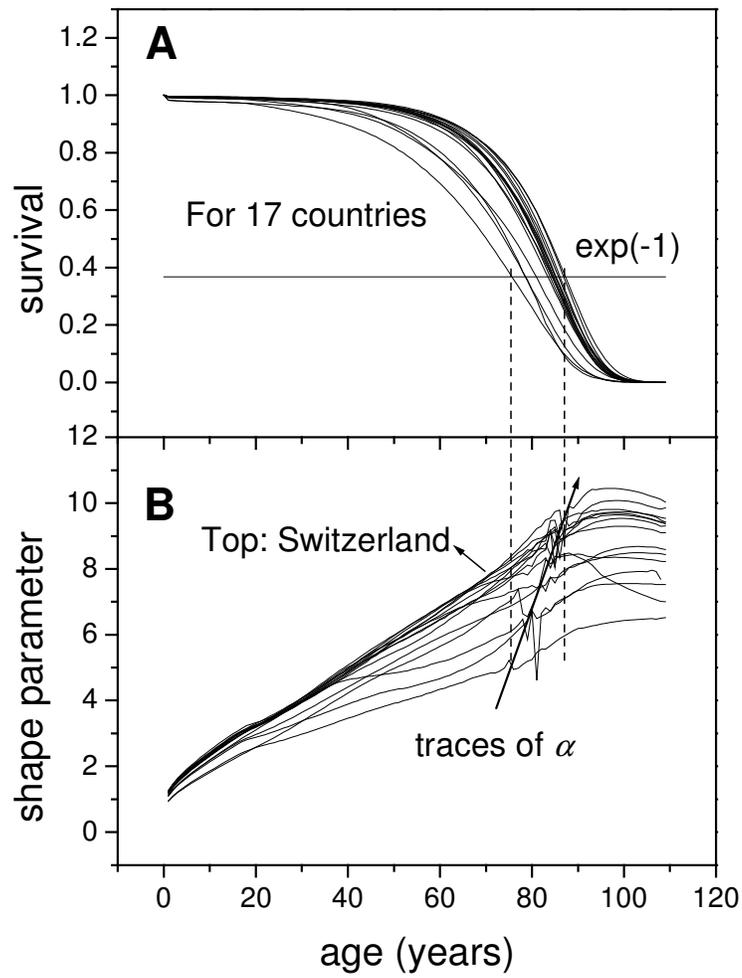

Fig. 2. Weon

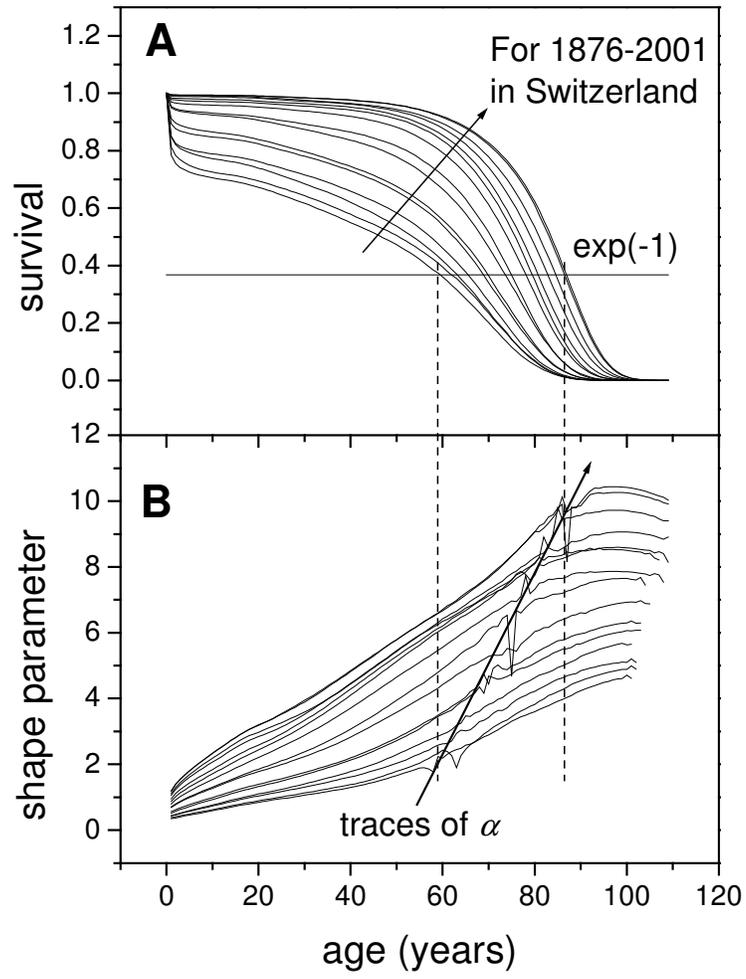

Fig. 3. Weon

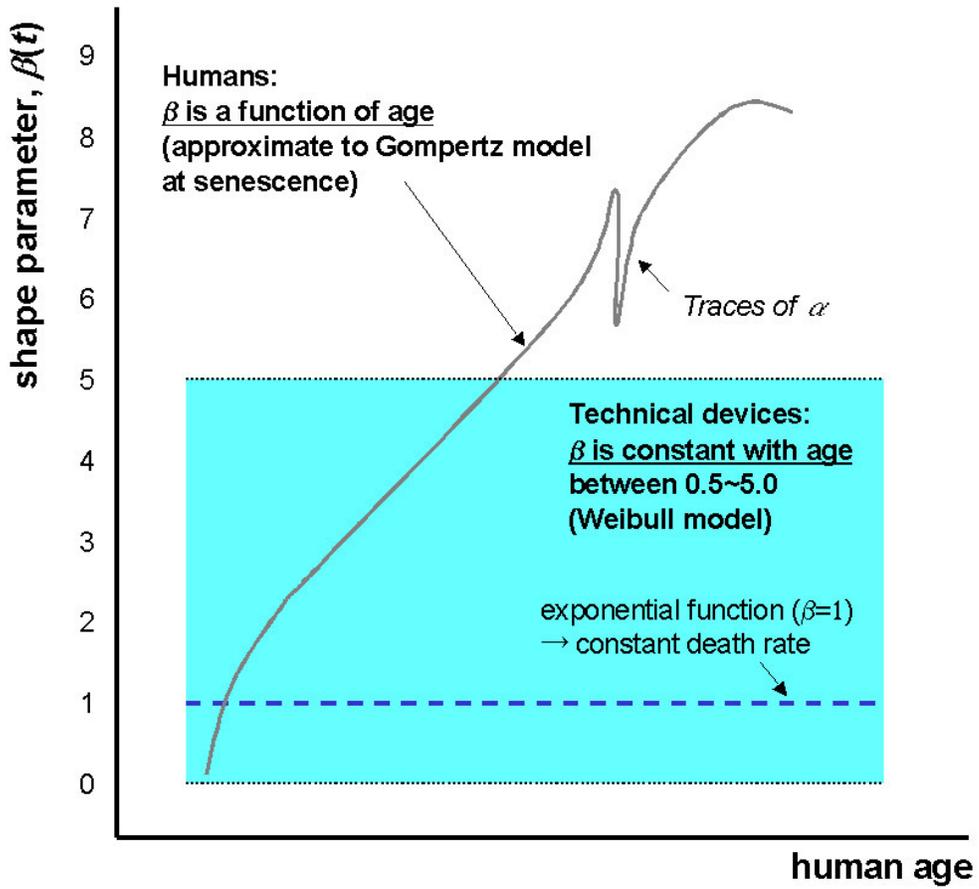

Fig. 4. Weon

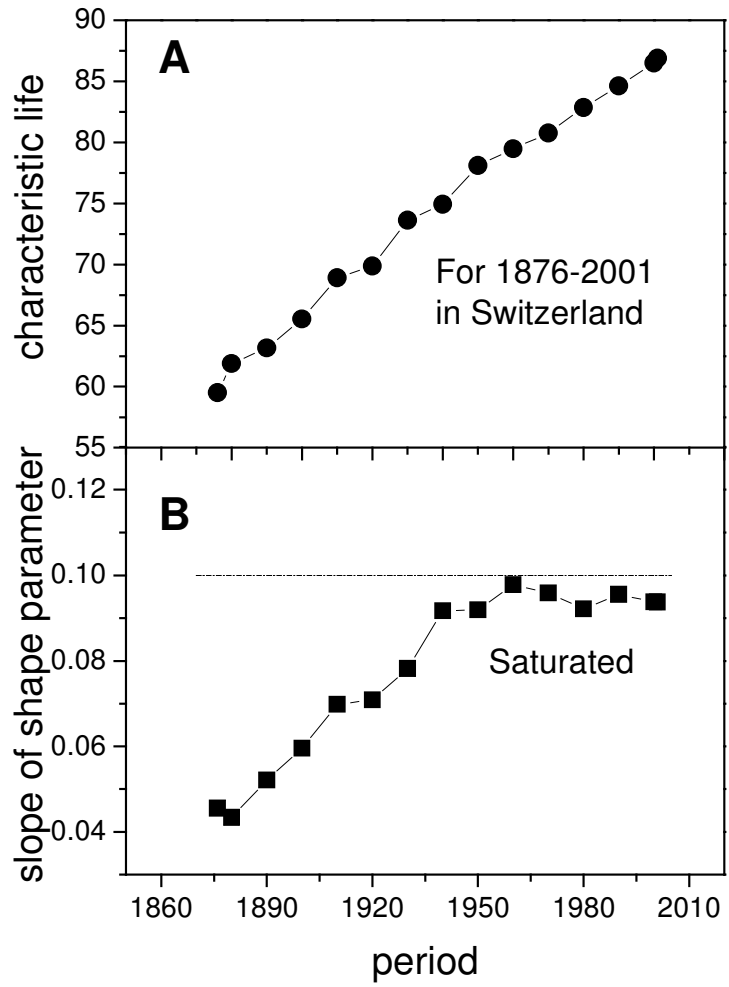

Fig. 5. Weon

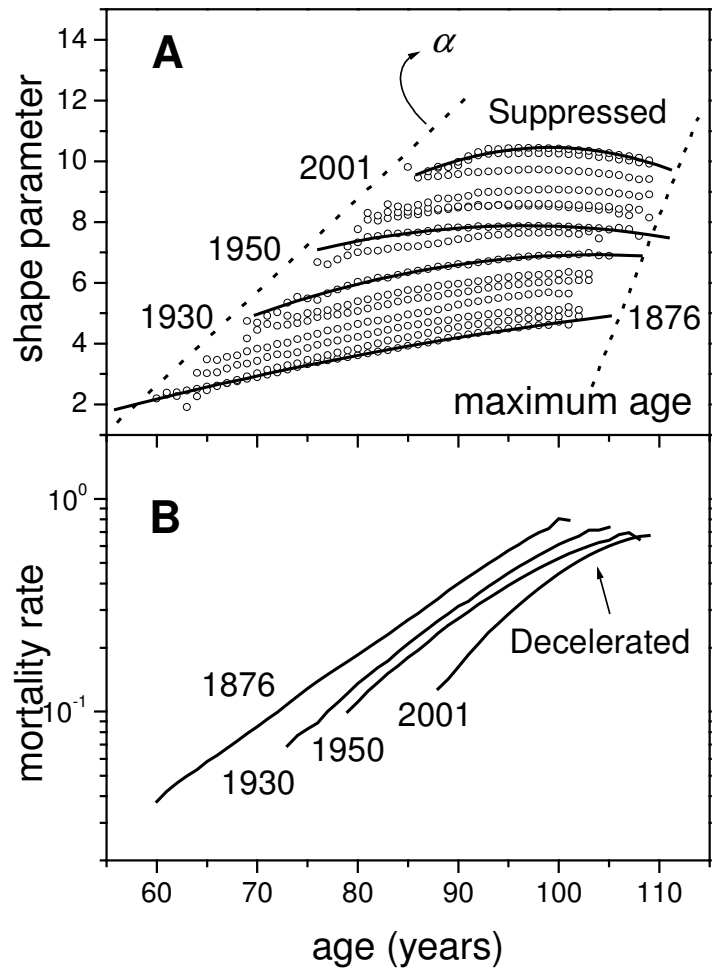

Fig. 6. Weon

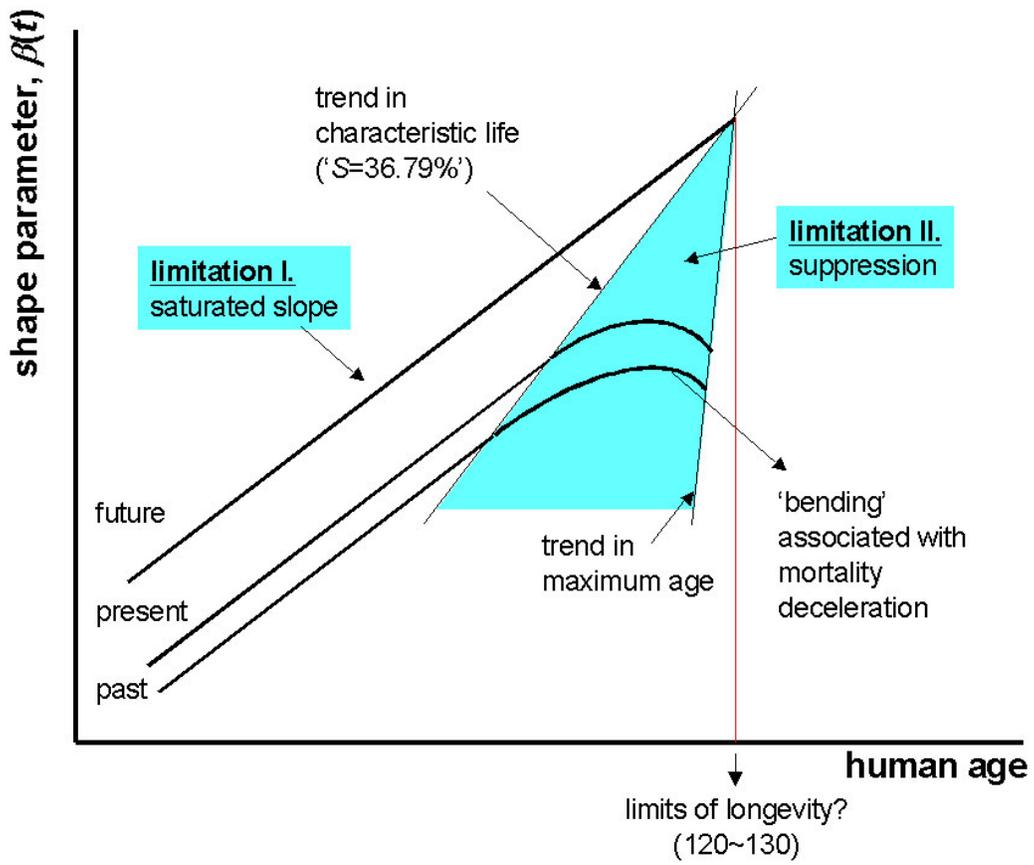

Fig. 7. Weon

Table 1. Characteristic life ($\alpha$) when survival=exp(-1)≈36.79% by country and period.

| Country (updated year) | $\alpha$ (years) | Period (Switzerland) | $\alpha$ (years) |
|---|---|---|---|
| Japan (1999) | 87.27 | 1876 | 59.51 |
| Switzerland (2001) | 86.88 | 1880 | 61.89 |
| Sweden (2001) | 86.05 | 1890 | 63.17 |
| Italy (1999) | 85.76 | 1900 | 65.53 |
| Norway (2000) | 85.25 | 1910 | 68.91 |
| Austria (1999) | 85.06 | 1920 | 69.86 |
| Canada (1996) | 85.02 | 1930 | 73.62 |
| Germany (1999) | 84.73 | 1940 | 74.93 |
| Finland (2000) | 84.64 | 1950 | 78.09 |
| Netherlands (1999) | 84.40 | 1960 | 79.47 |
| USA (1999) | 84.36 | 1970 | 80.75 |
| England & Wales (1998) | 83.94 | 1980 | 82.85 |
| Denmark (2000) | 83.47 | 1990 | 84.62 |
| Lithuania (2001) | 80.89 | 2000 | 86.49 |
| Hungary (1999) | 78.59 | 2001 | 86.88 |
| Bulgaria (1997) | 78.54 | | |
| Russia (1999) | 75.63 | | |

**Appendix: Resistance to ageing**

This is an appendix about a consideration of the question derived from the interesting finding that the shape parameter for humans is a function of age, not a constant value. The Weibull survival function for technical devices has a constant shape parameter, whereas the survival function for humans has an age-dependent shape parameter as follows,

$S = \exp(-(t/\alpha)^{\beta})$ --- technical devices

$S = \exp(-(t/\alpha)^{\beta(t)})$ --- humans

Consider the ageing processes, firstly following the Weibull survival function with a constant shape parameter, and then resisting to ageing by increasing the shape parameter. Assume that any limit factor fixes the characteristic life (Fig. A1). In this case, we can calculate a variation of duration of life by the shape parameter as follows,

$l = \alpha\delta = \alpha(-\ln S)^{1/\beta}$

where $\delta$ is introduced for convenience and called '*deviation parameter*' in this paper, and where $l$ means a specific duration of life when the survival ($S$) becomes a specific value. At the same value of the survival ($S$), the $l$ will be deviated according to the value of shape parameter ($\beta$). The deviation parameter ($\delta$) can indicate the deviation of the $l$ by the $\beta$. As the specific survival, we define a *half-characteristic survival* ($S_{hf}$) as the average between the origin (*S*=1) and the *characteristic survival* (*S$_c$*=exp(-1)) (corresponding to the characteristic life) as follows,

$$S_{hf} = \frac{1+\exp(-1)}{2} \approx 68.39\%$$

Empirically we know that this half-characteristic survival (*S$_{hf}$*) can show well the derivative of the *l* by the *β*. When we consider the resistance to ageing as '*a struggle to extend the duration of life against ageing (deterioration) by increasing the shape parameter*', we then evaluate quantitatively the resistance to ageing through the derivative of the *l* by the *β*, 'd*l*/d*β*'. That is, supposing that humans (biological systems) could resist to ageing to extend the *l* by increasing the *β*, the derivative 'd*l*/d*β*' would be a measure of the resistance to ageing.

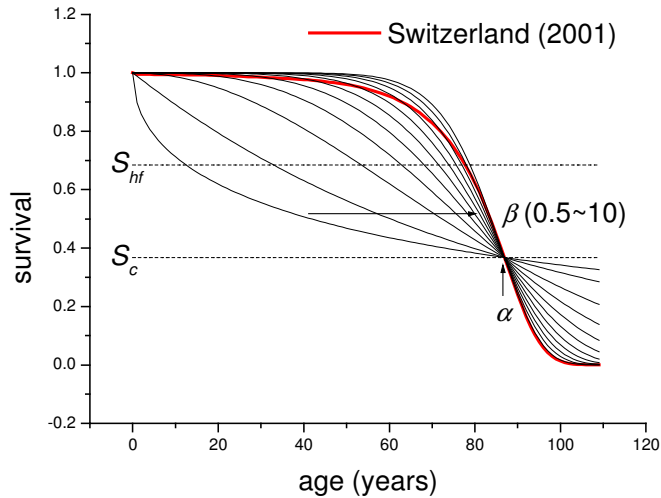

Fig. A1. Resistance to ageing by increasing the shape parameter.

The derivative 'd*l*/d*β*' can be defined at a specific survival and for a fixed characteristic life as follows,

$$\frac{dl}{d\beta} = \alpha(\frac{d\delta}{d\beta}) = \alpha(-\frac{\delta}{\beta}\ln\delta)$$

$$c = -\ln S = \delta^{\beta}$$
$$\ln c = \beta \ln \delta$$
$$\frac{1}{c}dc = \frac{\beta}{\delta}d\delta + \ln\delta d\beta$$
$$dc = 0 \text{ (at a specific survival)}$$
$$\therefore \frac{d\delta}{d\beta} = -\frac{\delta}{\beta}\ln\delta$$
$$\therefore \frac{dl}{d\beta} = \alpha(\frac{d\delta}{d\beta}) = \alpha(-\frac{\delta}{\beta}\ln\delta) \text{ (for a fixed characteristic life)}$$

We can simulate the derivative 'd*δ*/d*β*' (or d*l*/d*β*) with the *β* for the $S_{hf}$ (Fig. A2),

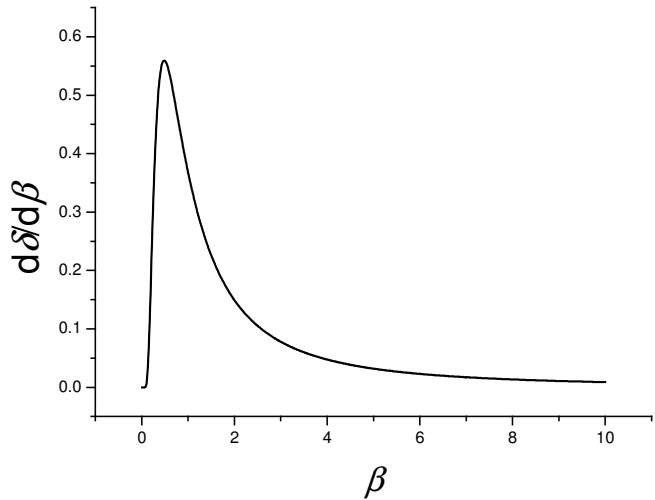

Fig. A2. Simulation of the derivative 'd*δ*/d*β*' with the *β* for the $S_{hf}$.

The shape parameter ($\beta$) is progressive with age and the resistance to ageing for humans (biological systems) can be described as a struggle to extend the duration of life by increasing the shape parameter, indicating that the resistance to ageing decreases with the $\beta$ ($\geq \sim 0.5$). We found out the shape parameter progression with age for human survival curves. The age-dependent shape parameter was changed from approximately 0.5 to 10 with age for the typical human survival curves. This feature is a great contrast to technical devices typically having a constant shape parameter. We think that this difference may be due to the resistance to ageing for humans (biological systems). This resistance is attributed to the *homeostasis* (resistance to change) or the *adaptation* of biological systems, which must have the homeostasis and the adaptation to maintain stability and to survive. Consequently, this hypothetical explanation shows that the shape parameter as a function of age for humans may be associated with the resistance to ageing.